\documentclass[aps,prl,twocolumn,showpacs,superscriptaddress,final]{revtex4-1}
\usepackage[final]{graphicx}
\usepackage{amsmath}
\usepackage{color}
\usepackage[outdir=./]{epstopdf}
\usepackage{float}
\usepackage[hidelinks = true]{hyperref}
\begin{document}

\title{Revealing the Coulomb interaction strength in a cuprate superconductor}

\author{S.-L. Yang}
\altaffiliation[current affiliations: ]{Kavli Institute at Cornell for Nanoscale Science; Laboratory of Atomic and Solid State Physics, Department of Physics; Department of Materials Science and Engineering. Cornell University, Ithaca, New York 14853, USA}
\affiliation{Stanford Institute for Materials and Energy Sciences, SLAC National Accelerator Laboratory, 2575 Sand Hill Road, Menlo Park, CA 94025, USA}
\affiliation{Geballe Laboratory for Advanced Materials, Departments of Physics and Applied Physics, Stanford University, Stanford, CA 94305, USA}
\author{J.~A. Sobota}
\affiliation{Stanford Institute for Materials and Energy Sciences, SLAC National Accelerator Laboratory, 2575 Sand Hill Road, Menlo Park, CA 94025, USA}
\affiliation{Advanced Light Source, Lawrence Berkeley National Laboratory, Berkeley, CA 94720, USA}
\author{Y. He}
\author{Y. Wang}
\author{D. Leuenberger}
\affiliation{Stanford Institute for Materials and Energy Sciences, SLAC National Accelerator Laboratory, 2575 Sand Hill Road, Menlo Park, CA 94025, USA}
\affiliation{Geballe Laboratory for Advanced Materials, Departments of Physics and Applied Physics, Stanford University, Stanford, CA 94305, USA}
\author{H. Soifer}
\affiliation{Stanford Institute for Materials and Energy Sciences, SLAC National Accelerator Laboratory, 2575 Sand Hill Road, Menlo Park, CA 94025, USA}
\author{M. Hashimoto}
\author{D.~H. Lu}
\affiliation{Stanford Synchrotron Radiation Lightsource, SLAC National Accelerator Laboratory, 2575 Sand Hill Road, Menlo Park, California 94025, USA}
\author{H. Eisaki}
\affiliation{Electronics and Photonics Research Institute, National Institute of Advanced Industrial Science and Technology, Tsukuba, Ibaraki 305-8558, Japan}
\author{B. Moritz}
\author{T.~P. Devereaux}
\author{P.~S. Kirchmann}
\email{kirchman@slac.stanford.edu}
\affiliation{Stanford Institute for Materials and Energy Sciences, SLAC National Accelerator Laboratory, 2575 Sand Hill Road, Menlo Park, CA 94025, USA}
\author{Z.-X. Shen}
\email{zxshen@stanford.edu}
\affiliation{Stanford Institute for Materials and Energy Sciences, SLAC National Accelerator Laboratory, 2575 Sand Hill Road, Menlo Park, CA 94025, USA}
\affiliation{Geballe Laboratory for Advanced Materials, Departments of Physics and Applied Physics, Stanford University, Stanford, CA 94305, USA}

\date{\today}

\begin{abstract}
We study optimally doped Bi$_{2}$Sr$_{2}$Ca$_{0.92}$Y$_{0.08}$Cu$_{2}$O$_{8+\delta}$ (Bi2212) using angle-resolved two-photon photoemission spectroscopy. Three spectral features are resolved near $1.5$, $2.7$, and $3.6$~eV above the Fermi level. By tuning the photon energy, we determine that the $2.7$~eV feature arises predominantly from unoccupied states. The $1.5$ and $3.6$~eV features reflect unoccupied states whose spectral intensities are strongly modulated by the corresponding occupied states. These unoccupied states are consistent with the prediction from a cluster perturbation theory based on the single-band Hubbard model. Through this comparison, a Coulomb interaction strength $U$ of $2.7$~eV is extracted. Our study complements equilibrium photoemission spectroscopy and provides a direct spectroscopic measurement of the unoccupied states in cuprates. The determined Coulomb $U$ indicates that the charge-transfer gap of optimally doped Bi2212 is $1.1$~eV.

\end{abstract}
\maketitle

\section{Introduction}

Governed by Fermi-Dirac statistics, electronic states above the Fermi level $E_{\rm F}$ are unoccupied at zero temperature~\cite{Ashcroft1976}. Studies of unoccupied states yield critical information about topological properties~\cite{Sobota2013} and symmetry-breaking orders~\cite{Hashimoto2010, Yang2008}. In particular, knowledge of unoccupied states is essential for determining the symmetry of a spectral gap, which encodes the origin of the corresponding order~\cite{Hashimoto2010, Yang2008}. For cuprate superconductors which host a complex interplay of competing orders~\cite{Keimer2015}, the ability to resolve unoccupied electronic states is particularly important.

A Mott insulating phase is a manifestation of strong correlation physics~\cite{Imada1998}. Due to Coulomb repulsions, half-filled electronic states are localized resulting in an insulating phase~\cite{Imada1998}. The hallmark of the Mott physics is the formation of lower Hubbard band (LHB) and upper Hubbard band (UHB), separated by the Coulomb interaction strength $U$. As the UHB is above $E_{\rm F}$ and unoccupied, an energy- and momentum-resolved characterization of UHB in cuprates has remained challenging.

Angle-resolved photoemission spectroscopy (ARPES) enables a direct measurement of the single-particle spectral function, which contains the information of electronic band structures and the underlying interactions~\cite{Damascelli2003,Shen1993,Lanzara2001}. However, the application of ARPES has been typically limited to the occupied part of the spectral function. Numerical techniques such as division by the Fermi-Dirac distribution have been used to reveal the states slightly above $E_{\rm F}$~\cite{Lee2007}, yet this method is confined to an energy range on the order of the sample temperature. A recent ARPES study on Bi-based cuprates identified features contributed by unbound states at $6$~eV above $E_{\rm F}$~\cite{Miller2015}. However, the key quantities of the strong correlation physics in cuprates - the energy scale of the UHB and the Coulomb interaction strength - remain underexplored.

Several techniques have studied the unoccupied electronic states in cuprates. Inverse photoemission spectroscopy (IPES) revealed unoccupied states from $0$ to $\sim 10$~eV above $E_{\rm F}$~\cite{Wagener1989,Wagener1990,Claessen1989,Drube1989,Bernhoff1990,Watanabe1991}.  However, IPES experiments are challenging due to the $10^5$-lower efficiency compared to ARPES~\cite{Johnson1985} and the $0.3\sim 1$~eV energy resolution~\cite{Claessen1989,Drube1989,Bernhoff1990}. X-ray absorption spectroscopy~\cite{Himpsel1988,Bianconi1992,Saini1996} and scanning tunneling spectroscopy (STS)~\cite{Ye2013} are also capable of characterizing the unoccupied states. Yet, these studies measure momentum-integrated density of states instead of momentum-resolved band structures. Two-photon photoemission (2PPE) enables the measurement of momentum-resolved unoccupied band structures with $<30$~meV energy resolution~\cite{Petek1997,Weinelt2002,Sobota2013,Sonoda2002,Sonoda2004,Gilbertson2014}. Pioneer 2PPE works on cuprates by Sonoda and Munakata revealed unoccupied states at the Brillouin zone center~\cite{Sonoda2002,Sonoda2004}. To further study the unoccupied band structure and the strong correlation physics, a momentum-resolved 2PPE study with a detailed comparison to theoretical calculations is needed.

Here we report a momentum-resolved 2PPE study on optimally doped Bi$_{2}$Sr$_{2}$Ca$_{0.92}$Y$_{0.08}$Cu$_{2}$O$_{8+\delta}$ (OP Bi2212, T$_{c}$ = $96$~K). Near the Brillouin zone center we resolve features near $1.5$, $2.7$, and $3.6$~eV above $E_{\rm F}$, denoted as $\alpha$, $\beta$, and $\gamma$, respectively. Tuning the photon energy from $4.5$ to $4.8$ eV, the binding energies of $\beta$ and $\gamma$ stay unchanged, whereas feature $\alpha$ becomes weak and unidentifiable. Comparison with the ARPES spectrum suggests that $\alpha$ as well as $\gamma$ correspond to unoccupied states whose spectral intensities are strongly modulated by the respective occupied states. Furthermore, we compare our results with calculations using the cluster perturbation theory (CPT), from which a Coulomb interaction strength $U$ of $2.7$~eV is extracted. Our study provides an important benchmark for studying correlation physics in cuprate superconductors.

\section{Methods}

Our optical setup is based on a regenerative amplifier system which typically outputs $1.5$~eV photons with $312$~kHz repetition rate, $<40$~fs pulse duration, and $\sim 6$~$\mu$J pulse energy. Two stages of nonlinear frequency conversions are employed: the first $\beta$-BaB$_{2}$O$_{4}$ (BBO) crystal yields the second harmonic; the second BBO sums the frequencies of the fundamental and the second harmonic. The third harmonic pulse duration is $<140$~fs. Its photon energy is tunable between $4.5$ and $4.8$~eV. The incident fluence for our measurements is $7$~$\mu$J.cm$^{-2}$. The $p$-polarized third harmonic is focused on optimally doped Bi2212 samples to conduct monochromatic 2PPE measurements. The photon polarization is orthogonal to the analyzer slit. For occupied-state studies, $6$~eV photons are generated by two stages of second harmonic generation from the $1.5$~eV laser. The energy resolution of $6$~eV ARPES is $22$~meV. We also take ARPES measurements using $22.7$~eV photons at the Stanford Synchrotron Radiation Lightsource, with a resolution of $6.5$~meV. The Bi2212 samples are grown using the traveling-solvent floating-zone technique~\cite{Eisaki2004}, and cleaved {\it in situ} under ultrahigh vacuum with a pressure $<7\times10^{-11}$~Torr. The measurement temperature is set at $20$~K.

Our theoretical calculation is based on a single-band Hubbard model solved by CPT~\cite{Senechal2000, Senechal2002, Wang2015}. Although CPT is an approximate method, we believe it is most suitable for the comparison with experimental data due to its continuous momentum resolution evaluated in a zero-temperature many-body wavefunction. We refer readers to Ref.~\cite{Wang2015} for a detailed implementation of the calculation.

\begin{figure}
\begin{center}
\includegraphics[width=\columnwidth]{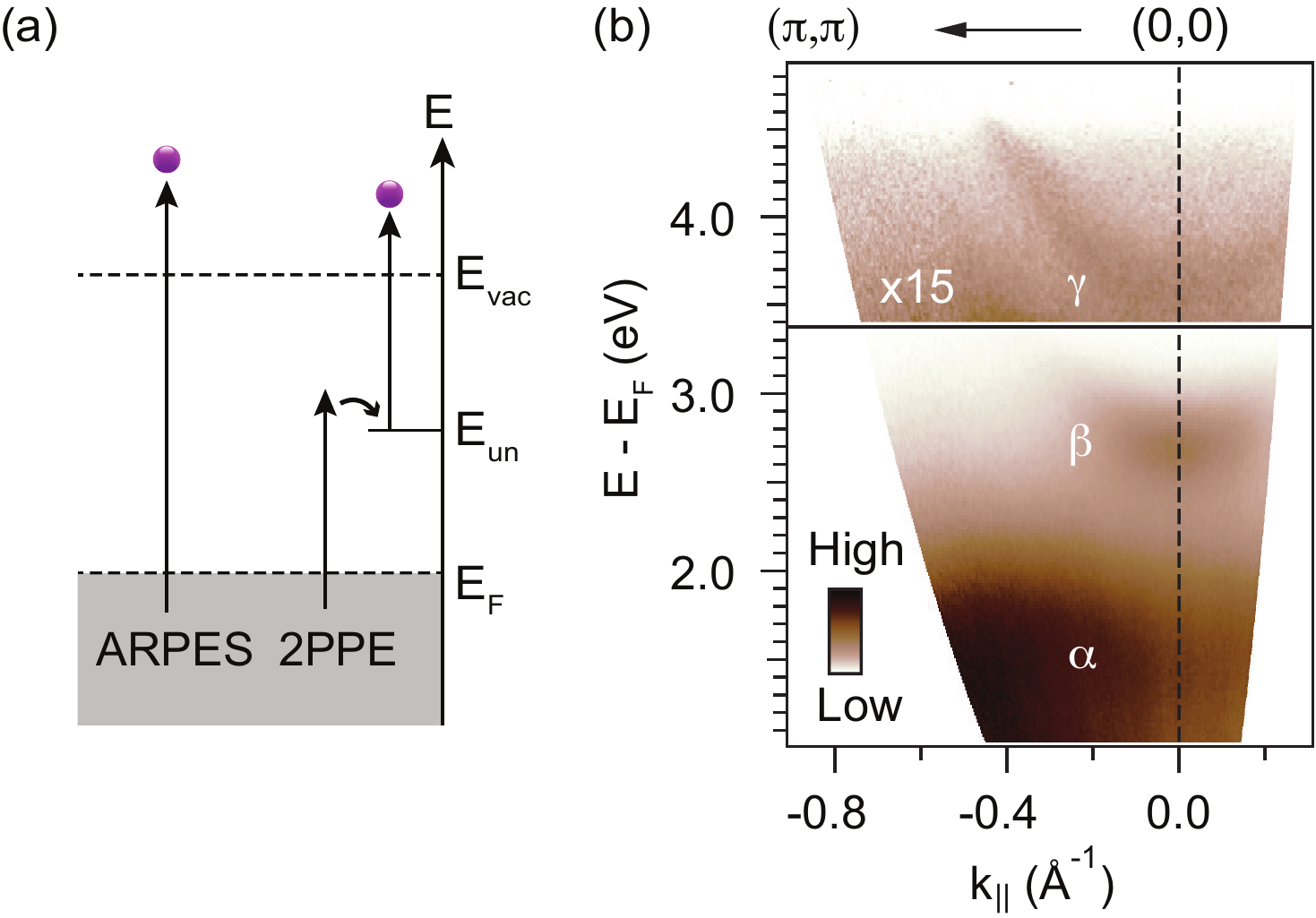}
\caption{Overview of the two-photon photoemission (2PPE) data on OP96 Bi2212 at 20 K. (a) Illustration of the ARPES and 2PPE processes. $E_{\rm F}$, $E_{\rm un}$, and $E_{\rm vac}$ are defined in the text. (b) 2PPE spectrum along the Brillouin zone diagonal. At the zone center, features $\alpha$, $\beta$, and $\gamma$ are identified near 1.5, 2.7, and 3.6 eV, respectively. The intensities in the energy range of 3.4$\sim$4.8 eV are magnified by a factor of $15$ to highlight the weak feature $\gamma$.}\label{Fig1}
\end{center}
\end{figure}

\section{Results}

We present an overview of the 2PPE spectrum using $4.5$~eV photons in Fig.~\ref{Fig1}. Figure~\ref{Fig1}(a) illustrates the one-photon excitation in ARPES and the two-photon excitation in 2PPE~\cite{Sobota2013}. For the latter, the first photon promotes electrons from occupied states below $E_{\rm F}$ to high-lying unoccupied states. Scattering processes can occur to populate the lower-energy unoccupied states at energy $E_{\rm un}$. These intermediate states are subsequently promoted by the second photon to final states above the vacuum level $E_{\rm vac}$. Throughout this work we follow the usual convention of ARPES experiments and discuss the binding energies of the intermediate states referenced to $E_{\rm F}$ on the detector. This defines the intermediate state energy scale~\cite{Sonoda2004}, which allows a consistent comparison between the occupied and unoccupied states. In Fig.~\ref{Fig1}(b) we display the 2PPE spectrum along the $(0,0)$-$(\pi,\pi)$ direction. At the zone center ($\Gamma$) we identify features near $1.5$~eV ($\alpha$), $2.7$~eV ($\beta$), and $3.6$~eV ($\gamma$). The observed features are consistent with previous 2PPE measurements at $\Gamma$~\cite{Sonoda2002,Sonoda2004}.

\begin{figure}
\begin{center}
\includegraphics[width=\columnwidth]{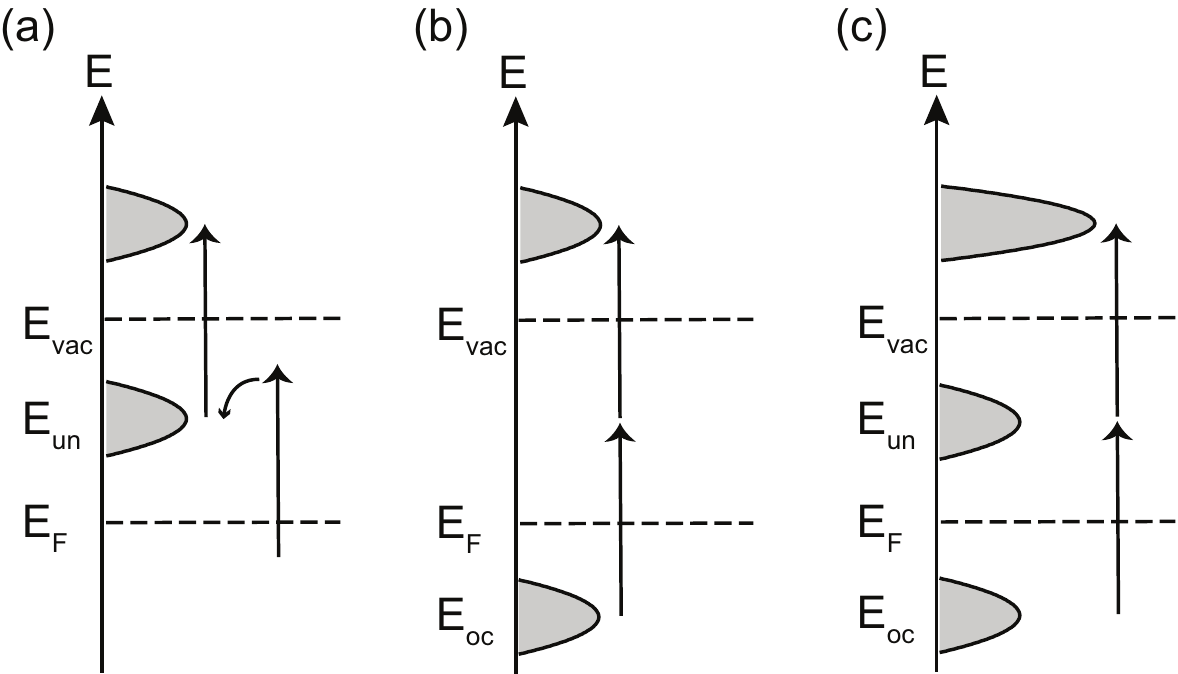}
\caption{Illustration of different two-photon excitation schemes. Notations are the same as in Fig.~\ref{Fig1}(a). (a) Unoccupied-state spectroscopy. The resolved binding energy is ($E_{\rm un} - E_{\rm F}$) and does not depend on the photon energy. (b) Occupied-state spectroscopy. The resolved binding energy is ($E_{\rm oc} + h\nu - E_{\rm F}$) and depends on the photon energy. (c) Resonant excitation of unoccupied states. The spectral intensity is enhanced with respect to the non-resonant schemes in (a) and (b).}\label{Fig2}
\end{center}
\end{figure}

\begin{figure*}
\begin{center}
\includegraphics[width=5.5in]{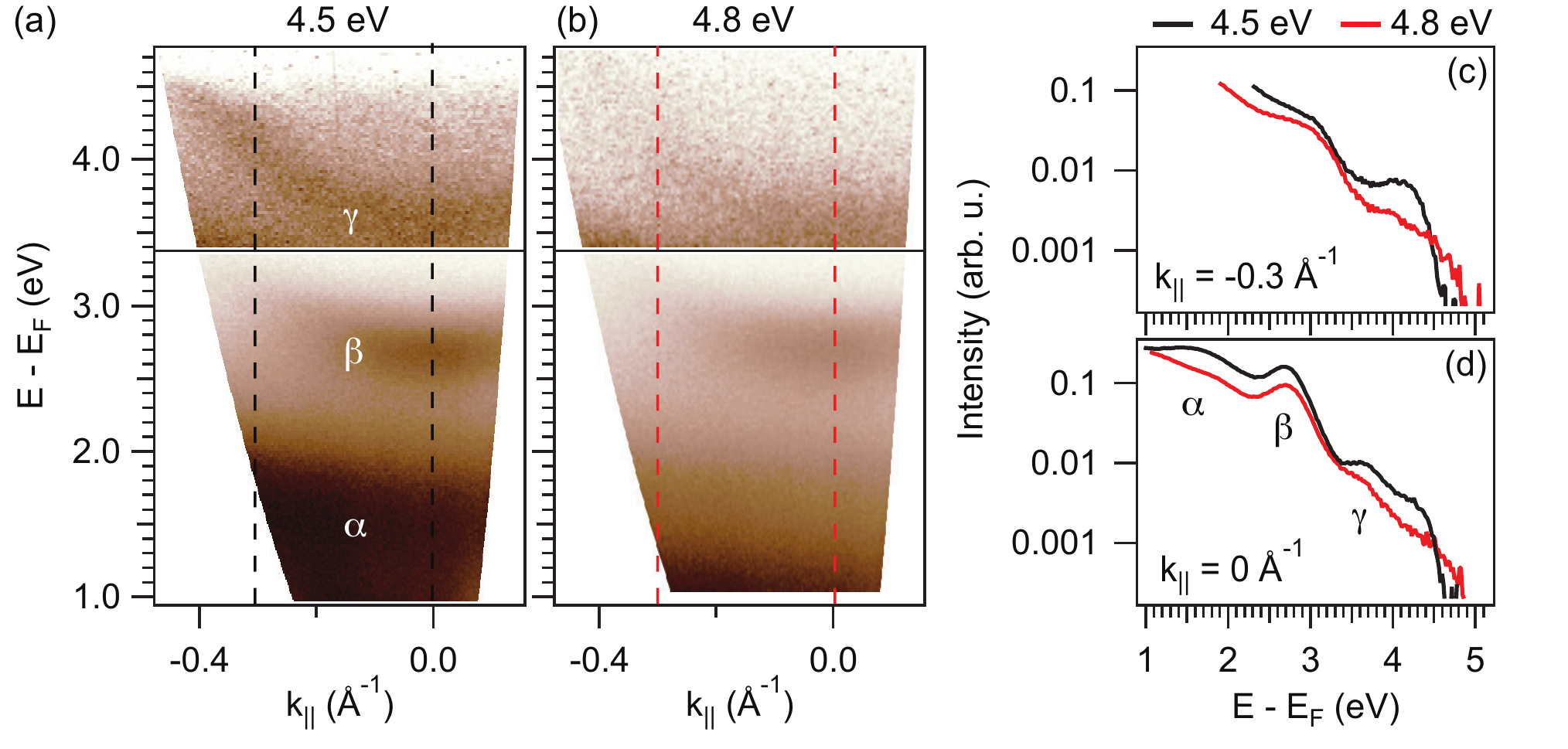}
\caption{Photon energy dependence of the 2PPE features. (a,~b) 2PPE spectra near the zone center using (a) $4.5$~eV and (b) $4.8$~eV photons. Both spectra are obtained with $9.5\times 10^{12}$~photons/(pulse.cm$^{2}$). (c,~d) Comparison of the EDCs at (c) $k_{||} = -0.3$~\AA$^{-1}$ and (d) $k_{||} = 0$~\AA$^{-1}$. Each EDC is obtained by integrating over a momentum window of $0.1$~\AA$^{-1}$. The cuts are indicated by dashed lines in panels (a) and (b).}\label{Fig3}
\end{center}
\end{figure*}

\begin{figure}
\begin{center}
\includegraphics[width=\columnwidth]{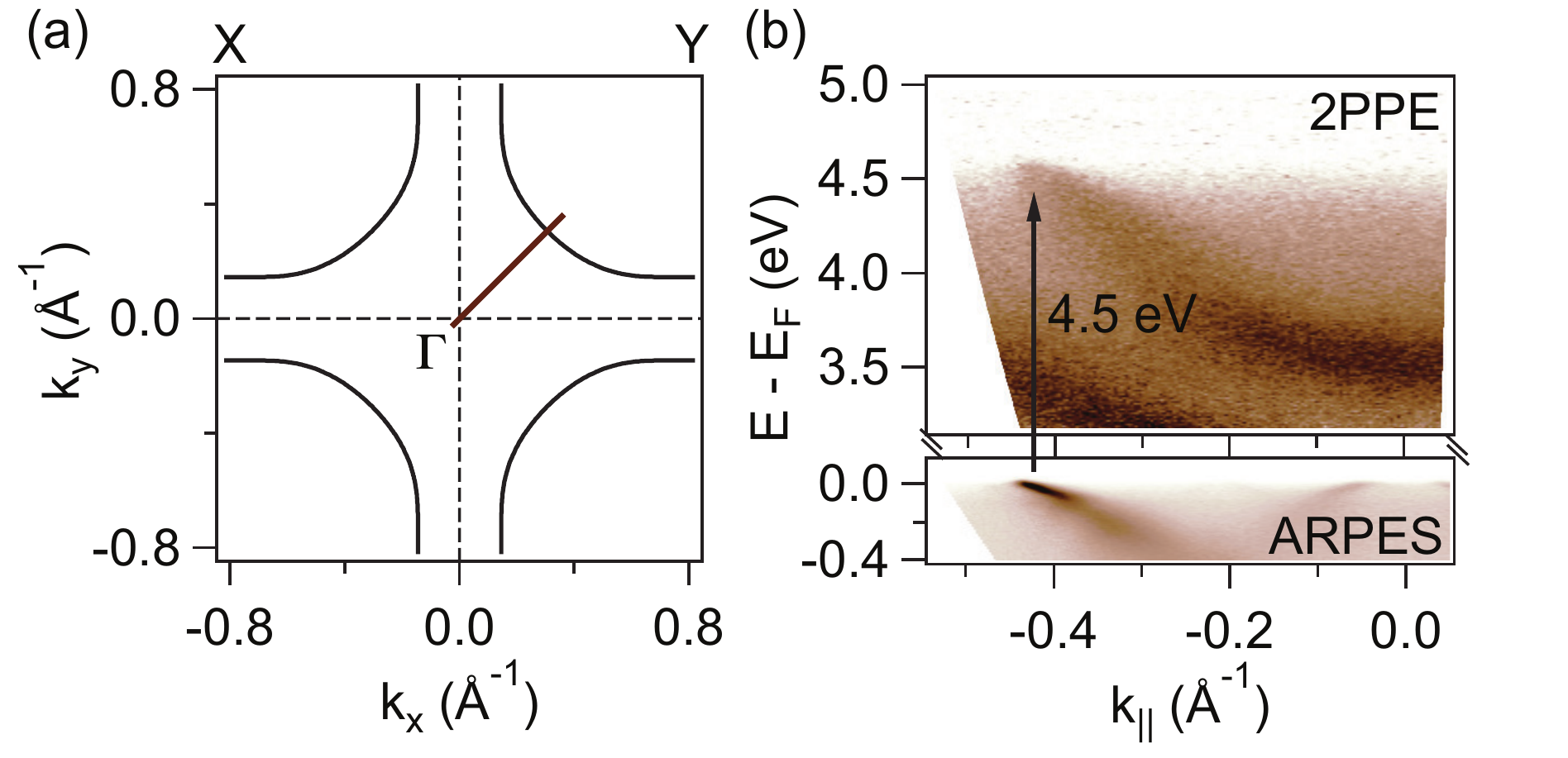}
\caption{Resonant 2PPE process for feature $\gamma$ using $4.5$~eV photons. (a) Illustration of the momentum-space trajectory (solid brown) for the spectra in (b). Overlaid on the graphs are the tight-binding Fermi surfaces (solid black)~\cite{Markiewicz2005}. (b) 2PPE spectrum of feature $\gamma$ using $4.5$~eV photons and ARPES spectrum of the occupied states near $E_{\rm F}$ using $6$~eV photons.}\label{Fig4}
\end{center}
\end{figure}

Importantly, 2PPE can be used to probe both the occupied and unoccupied states~\cite{Sonoda2002,Sonoda2004,Sobota2013}. Figure~\ref{Fig2}(a) illustrates the ideal unoccupied-state spectroscopy where the 2PPE spectrum is predominantly determined by unoccupied states. In this case, the resolved binding energy is ($E_{\rm un} - E_{\rm F}$) and does not depend on the photon energy. Meanwhile, a distinct 2PPE process in Fig.~\ref{Fig2}(b) shows that occupied states at energy $E_{\rm oc}$ can be photoemitted by a direct two-photon process. The binding energy of the virtual intermediate state increases linearly with photon energies. Moreover, a resonant excitation scheme can occur when an occupied state is projected to an unoccupied state by the first photon (Fig.~\ref{Fig2}(c)). In this case, the spectral intensity is much enhanced compared to the non-resonant cases in Fig.~\ref{Fig2}(a) and (b).

To distinguish between different excitation scenarios, we perform a photon energy dependent study on the 2PPE spectrum (Fig.~\ref{Fig3}). Spectra in Fig.~\ref{Fig3}(a) and (b) are obtained with $4.5$ and $4.8$~eV photons, respectively. The incident beam flux is maintained at $9.5\times 10^{12}$ photons/(pulse.cm$^{2}$). We compare energy distribution curves (EDCs) taken at constant momentum points in Fig.~\ref{Fig3}(c) and (d). At $k_{||} = 0$~\AA$^{-1}$, features $\beta$ and $\gamma$ display negligible shifts when tuning the photon energy, which indicates that they correspond to unoccupied states. Intriguingly, using $4.5$~eV photons the spectral intensity of feature $\gamma$ at $k_{||} = -0.3$~\AA$^{-1}$ is significantly higher than that using $4.8$~eV photons (Fig.~\ref{Fig3}(c)). The spectral peak of feature $\alpha$ using $4.8$~eV photons becomes unidentifiable. These observations suggest that features $\alpha$ and $\gamma$ are substantially influenced by their corresponding initial states~\cite{Sobota2013}.

To examine the optical excitation for feature $\gamma$, we compare the 2PPE spectrum using $4.5$~eV photons with the ARPES spectrum using $6$~eV photons (Fig.~\ref{Fig4}). In Fig.~\ref{Fig4}(a) we plot the Fermi surface calculated by a tight-binding model~\cite{Markiewicz2005}. The momentum trajectory along $(0,0)$-$(\pi,\pi)$ intercepts the Fermi surface, resulting in the occupied-state dispersion measured by $6$~eV ARPES, as shown in the lower panel of Fig.~\ref{Fig4}(b). Photoexcitations promote this occupied state to $4.5$~eV above $E_{\rm F}$, leading to the dispersive feature in the 2PPE spectrum near $-0.4$~\AA$^{-1}$. This resonant excitation explains the enhancement in spectral intensities of feature $\gamma$ using $4.5$~eV photons. In the ARPES spectrum we also observe band structures near the zone center induced by the incommensurate modulation of the BiO planes along the crystallographic {\it b} axis~\cite{Mans2006}. It is challenging to determine whether the same effect is observed in the 2PPE spectrum due to the strong diffuse background.

To investigate the optical excitation for feature $\alpha$, we compare the 2PPE spectrum using $4.5$~eV photons with the valence-band ARPES spectrum using $22.7$~eV photons at Stanford Synchrotron Radiation Lightsource. We notice that feature $\alpha$ is almost non-dispersive across the entire Brillouin zone, which resembles the characteristics of localized non-bonding states. As shown in Fig.~\ref{Fig5}, by shifting the ARPES spectrum $4.5$~eV upwards, a clear correspondence is established between feature $\alpha'$ on the ARPES spectrum and feature $\alpha$ on the 2PPE spectrum. Previous ARPES studies have identified feature $\alpha'$ as a non-bonding oxygen $2p$ state~\cite{He2016}, which explains the non-dispersive character of feature $\alpha$. Therefore, Fig.~\ref{Fig5} demonstrates that $\alpha$ originates mostly from the non-bonding oxygen $2p$ state.

Our interpretation of feature $\alpha$ is different from that in a previous 2PPE study~\cite{Sonoda2004}. Ref.~\cite{Sonoda2004} attributed feature $\alpha$ purely to the UHB, which is an unoccupied state. However, the UHB is highly dispersive across the Brillouin zone~\cite{Wang2015,Moritz2009,Kusko2002}, which is inconsistent with our observation on feature $\alpha$. We emphasize that the modulation in intensity due to initial-state dispersions is key to understanding the origin of feature $\alpha$.

\section{Discussion}

Various techniques have been used to study the origins of the unoccupied states in cuprates. IPES studies in the early 1990s observed features near $2.9$ and $4$~eV~\cite{Wagener1989,Wagener1990,Claessen1989,Drube1989,Bernhoff1990,Watanabe1991}, which likely correspond to features $\beta$ and $\gamma$ in this work. Influenced by the band structure calculations available by then~\cite{Krakauer1988,Massidda1988}, most IPES studies attributed features $\beta$ and $\gamma$ to BiO bands. However, several issues have been noticed with this assignment. First, the band structure calculations~\cite{Krakauer1988,Massidda1988} are based on the local density approximation, which is questionable for strongly correlated materials such as cuprates. Second, as pointed out by Ref.~\cite{Claessen1989} the observed dispersions of features $\beta$ and $\gamma$ are vastly different from the predicted dispersions of the BiO bands~\cite{Krakauer1988,Massidda1988}.

Previous 2PPE studies conducted polarization dependence study to investigate the origins of the unoccupied states~\cite{Sonoda2002,Sonoda2004}. It was shown that $\beta$ and $\gamma$ disappear when photons are $s$-polarized, yet $\alpha$ survives for both $p$- and $s$-polarized photons. Accordingly, they concluded that $\beta$ and $\gamma$ have out-of-plane characters consistent with the Cu $d_{z^{2}}$ orbital, and that $\alpha$ has in-plane characters consistent with the Cu $d_{x^2-y^2}$ orbital. This interpretation assigns the unoccupied states to states in the CuO$_{2}$ layers where the many-body Mott physics occurs.

\begin{figure}
\begin{center}
\includegraphics[width=\columnwidth]{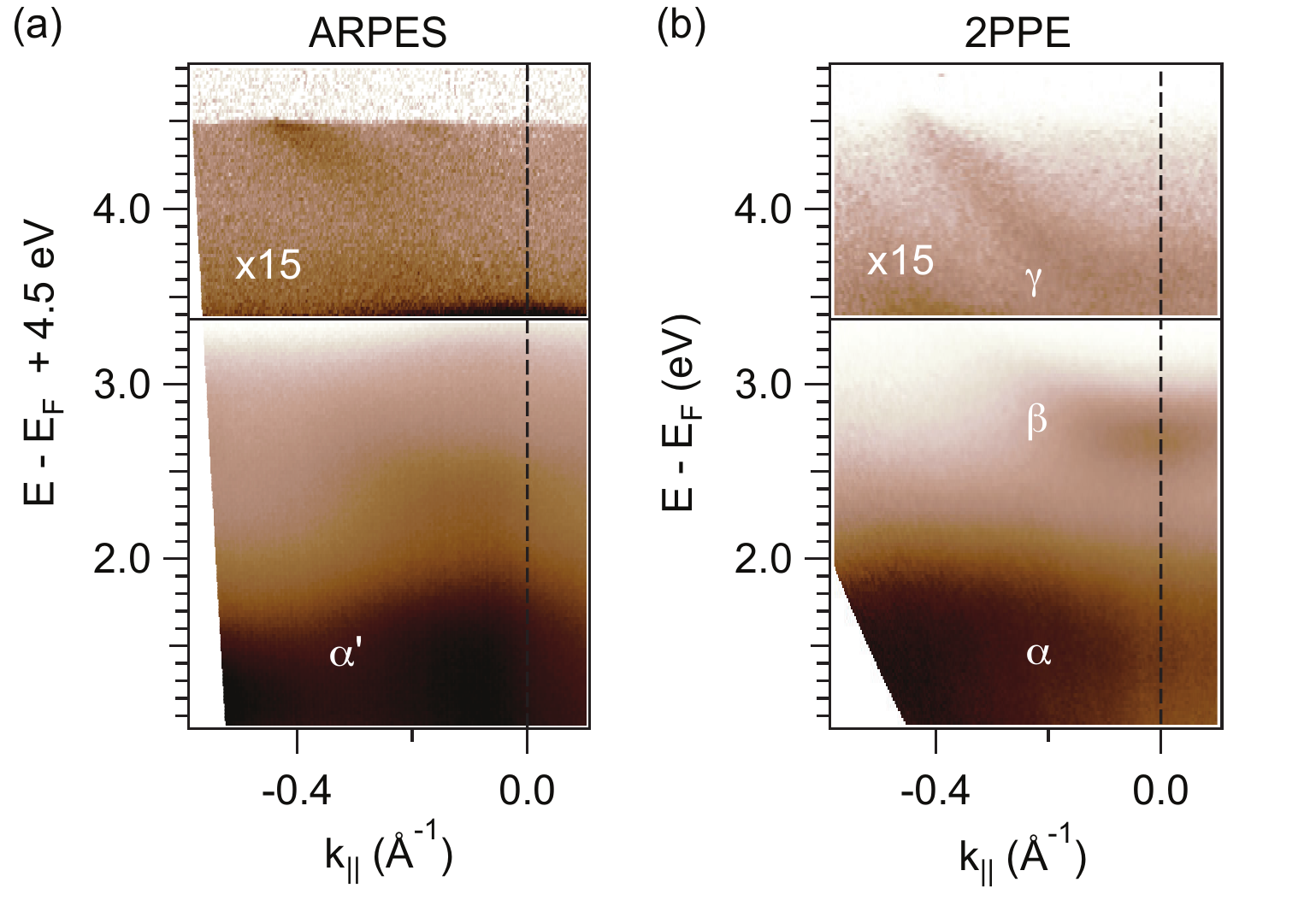}
\caption{Influence of initial states on feature $\alpha$. (a) ARPES spectrum of OP96 Bi2212 along the zone diagonal using $22.7$~eV photons. The energy axis is offset by $4.5$~eV to be compared with the 2PPE spectrum. (b) 2PPE spectrum on the same sample using $4.5$~eV photons.}\label{Fig5}
\end{center}
\end{figure}

\begin{figure*}[htp]
\begin{center}
\includegraphics[width=6in]{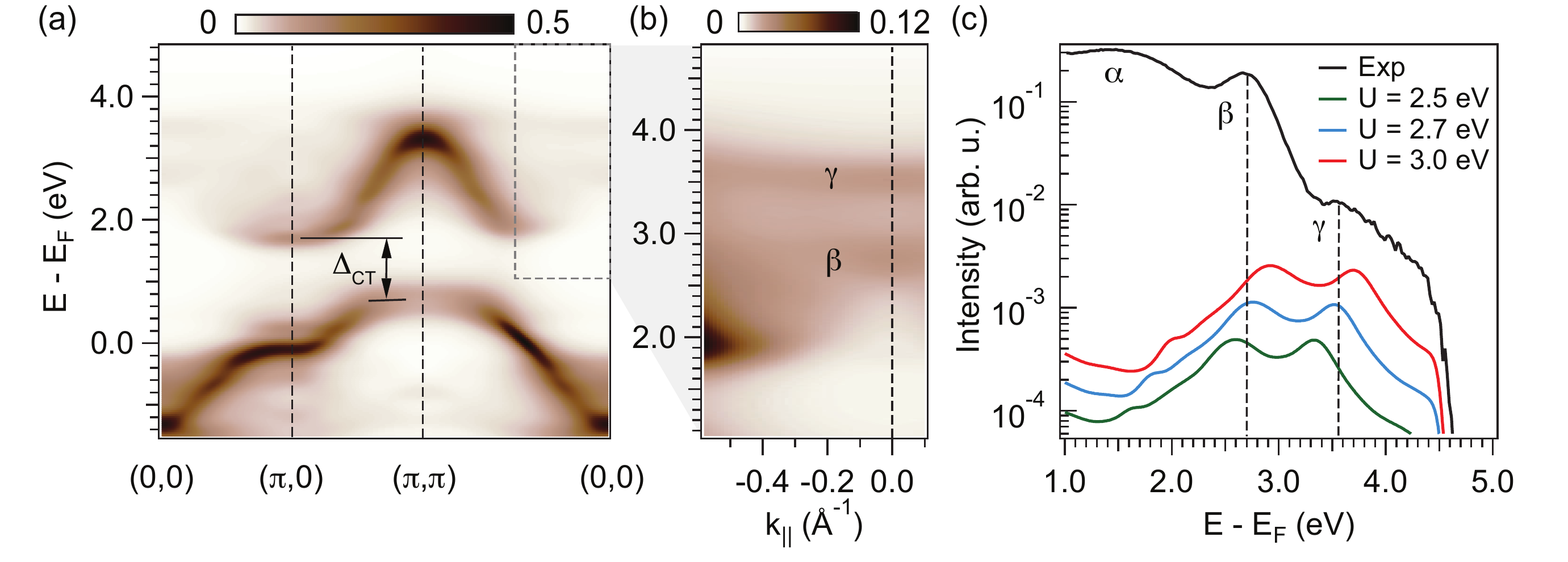}
\caption{Calculated spectrum using cluster perturbation theory. (a) Theoretical single-particle spectral function calculated by CPT for optimal doping and $U = 2.7$~eV. The charge-transfer gap $\Delta_{\rm CT} \sim 1.1$~eV is marked on the spectrum. (b) Theoretical unoccupied spectrum in the experimentally measured momentum and energy range. The corresponding color scales are indicated on top of panels (a) and (b). (c) Comparison of an experimental EDC taken at the zone center and the theoretical counterparts for various $U$ values. EDCs are offset on a log scale for clear demonstration. The theoretical EDCs are multiplied by a Fermi-Dirac function with the Fermi edge at $4.5$~eV above $E_{\rm F}$, which is the cutoff energy for 2PPE. The best match is obtained for $U = 2.7$~eV.}\label{Fig_CPT}
\end{center}
\end{figure*}

To obtain further understanding of the Mott physics, we compare our experimental results with a CPT calculation based on the single-band Hubbard model, which exclusively captures the low-energy Mott physics in CuO$_{2}$ planes~\cite{Wang2015}. The Hubbard Hamiltonian is comprised of a nearest (next nearest) neighbor hopping term parametrized by energy t (t$'$), and a Coulomb repulsion term parametrized by the interaction strength $U$. For cuprate superconductors, this Coulomb $U$ corresponds to the Cu-O charge transfer gap $\Delta_{\rm CT}$~\cite{Zaanen1985,Damascelli2003,Lee2006,Xiang2009}. We include only the Zhang-Rice singlet band~\cite{Zhang1988} in the single-band Hubbard model, and solve for the spectral function $A(k,\omega)$. Figure~\ref{Fig_CPT}(a) demonstrates the calculated spectrum corresponding to optimal doping and $U=2.7$~eV. Here we adopt $t=0.4$~eV determined from previous ARPES experiments~\cite{Xie2007}, and $t' = -0.3t$. It is worth noting that the UHB is comprised of fine features corresponding to different electron hopping mechanisms in the energy range of $2$ to $4$~eV~\cite{Wang2015}.

To compare the theoretical results with the experimental data, we emphasize that the entire feature $\alpha$ and feature $\gamma$ at $|k_{||}|>0.2$~\AA$^{-1}$ are strongly modulated by the occupied states, and hence should not be compared directly to the pure unoccupied states obtained by theory. Restraining our discussion to features $\beta$ and $\gamma$ near the zone center, we identify the two features on the theoretical spectrum as shown in Fig.~\ref{Fig_CPT}(b). We further plot the experimental EDC at $\Gamma$, and compare it to theoretical EDCs for a series of $U$ values (Fig.~\ref{Fig_CPT}(c)). Although the spectral shapes of $\beta$ and $\gamma$ depend on matrix elements and inelastic scattering processes~\cite{Sobota2012}, the peak positions can be utilized for a quantitative comparison. Varying $U$ between $2.4$ and $3.2$~eV with an increment of $0.08$~eV, we determine that the optimal matching between theory and experiment is achieved when $U = 2.7$~eV.

The comparison between CPT calculations and 2PPE results suggests that features $\beta$ and $\gamma$ at the zone center both belong to the UHB. Specifically, these features reflect the inter- and intra-sublattice electron motions~\cite{Wang2015}. We emphasize that there can be additional contributions from different origins. For features $\beta$ and $\gamma$, contributions from the $d_{z^{2}}$ orbital cannot be excluded~\cite{Sonoda2002,Sonoda2004}. For feature $\gamma$, the binding energy with respect to $E_{\rm vac}$ is close to that of the $n=1$ image potential state (IPS)~\cite{Hofer1997}. However, it is not readily evident in the 4.8 eV data (Fig.~\ref{Fig3}(b)) that feature $\gamma$ possesses a free-electron-like dispersion expected for an IPS. Hence a contribution of the IPS to feature $\gamma$ is unlikely but cannot be excluded.

There are a few important differences between theory and experiment. First, the theoretical spectrum contains a sharp feature near $2$~eV and $-0.5$~\AA$^{-1}$ which is not resolved experimentally. In 2PPE, this sharp feature can be overwhelmed by the strong modulation due to the occupied state $\alpha'$ (Fig.~\ref{Fig5}(a)). Second, the theoretical $\beta$ and $\gamma$ features in Fig.~\ref{Fig_CPT}(b) are rather non-dispersive. To avoid complications due to occupied states, we compare the theoretical features to the 2PPE results obtained in a non-resonant excitation regime (Fig.~\ref{Fig3}(b)). Here the spectral intensities of $\beta$ and $\gamma$ quickly decrease as a function of momentum away from $\Gamma$, which makes it challenging to determine the exact band dispersions. Further investigations are needed to quantify the experimental dispersions of $\beta$ and $\gamma$.

Nevertheless, the overall agreement between the CPT calculation and our 2PPE experiment has important implications. Momentum-resolved 2PPE lets us identify the UHB at the zone center, and furthermore the Coulomb interaction strength $U$. The Coulomb $U$ represents the energy cost forming a doubly-occupied state on a Cu site (doublon)~\cite{Xiang2009}, and was determined by earlier experiments which did not resolve the UHB~\cite{Shen1987}. Our study showcases a modern method to directly unveil the UHB and deduce the Coulomb $U$, which provides the basis for theoretical modeling of superconductivity and magnetism based on the single-band Hubbard model.

Taking into account the quasiparticle bandwidth $4t$~\cite{Xie2007}, our measurement suggests a charge-transfer gap $\Delta_{\rm CT} \sim U-4t = 1.1$~eV. This is a factor of two smaller than the counterparts in undoped La$_{2}$CuO$_{4}$~\cite{Falck1992}, Ca$_{2}$CuO$_{2}$Cl$_{2}$~\cite{Ye2013}, and Bi2201~\cite{Cai2016}. On the other hand, our result is consistent with gap values reported by optical spectroscopies on doped Bi2212~\cite{Itoh1999} and STS on undoped Bi2212~\cite{Ruan2016}. These comparisons suggest that $\Delta_{\rm CT}$ varies substantially between different cuprate families. A recent STS study~\cite{Ruan2016} discovered an anticorrelation between $\Delta_{\rm CT}$ in the parent compound and the maximum superconducting transition temperature $T_{\rm c}$ upon doping. This indicates a direct connection between electronic correlations and the superconducting pairing mechanism.

Interestingly, our results provide a new perspective to understand the chemical potential puzzle in the cuprate literature, where people have found a chemical potential shift $< 1$~eV when tuning from electron doping to hole doping~\cite{Ikeda2010}. This shift is supposed to match $\Delta_{\rm CT}$, yet the experimental value is much smaller than the conventional $\Delta_{\rm CT}$ of $\sim 2$~eV~\cite{Falck1992,Ye2013, Waku2004,Cai2016}. Our results show that in hole-doped Bi2212 $\Delta_{\rm CT}$ is as small as $1$~eV, which suggests that this apparent discrepancy in the literature is due to comparison across different material families with different magnitudes of $\Delta_{\rm CT}$. Notably, a careful analysis of the photoemission and optical spectroscopy data on electron-doped Nd$_{2}$CuO$_{4}$ yields a gap of $\sim 0.5$~eV~\cite{Xiang2009}. These values would be consistent with a chemical potential shift $<1$~eV when tuning from electron doping to hole doping. Future 2PPE experiments on electron-doped cuprates are clearly needed to verify this picture.

\section{Conclusion}

Our momentum-resolved 2PPE measurement characterizes the unoccupied band structure for optimally doped Bi2212. By tuning the photon energy, we identify an unoccupied state near $2.7$~eV above $E_{\rm F}$. Two other features near $1.5$ and $3.6$~eV reflect unoccupied states strongly modulated by occupied-state dispersions. These results are compared with the UHB spectrum calculated by CPT, which yields a Coulomb interaction strength $U$ of $2.7$~eV and a charge-transfer gap of $1.1$~eV. Notably, our study provides a clean method to characterize the Coulomb repulsion for doped Mott insulators. Our technique is advantageous compared to optical measurements which are complicated by the emergence of Drude peaks for finite doping~\cite{Terasaki1990, Itoh1999}. If the 2PPE measurement conditions are further optimized, it is conceivable that the full unoccupied band structure can be determined unambiguously. In the study of advanced materials such as cuprates~\cite{Damascelli2003} or iridates~\cite{Kim2014}, obtaining the full unoccupied band structure can determine the gap symmetries corresponding to various symmetry-breaking orders~\cite{Hashimoto2010, Yang2008}, which will be key to understanding the complex phase diagrams.

\begin{acknowledgments}
{\bf Acknowledgments} This work was supported by the U.S. Department of Energy, Office of Science, Basic Energy Sciences, Materials Sciences and Engineering Division under contract DE-AC02-76SF00515. S.-L.Y. and Y.W. acknowledge support by the Stanford Graduate Fellowship. S.-L.Y. is also supported by the Kavli Postdoctoral Fellowship at Cornell University. J.A.S. is in part supported by the Gordon and Betty Moore Foundations EPiQS Initiative through Grant GBMF4546. D.L. acknowledges partial support by the Swiss National Science Foundation under fellowship P300P2151328. H.S. acknowledges support from the Fulbright Scholar Program. Stanford Synchrotron Radiation Lightsource is operated by the U.S. Department of Energy, Office of Science, Office of Basic Energy Sciences.
\end{acknowledgments}

\bibliography{YangSL_Bi2212_2PPE_refs}

\end{document}